\documentstyle[12pt]{article}
\newcommand\be{\begin{equation}}
\newcommand\ee{\end{equation}}
\newcommand\bea{\begin{eqnarray}}
\newcommand\eea{\end{eqnarray}}

\newcommand{\fatalpha}{{\bf \alpha \kern -0.44em \alpha}}
\newcommand{\fatsigma}{{\bf \sigma \kern -0.54em \sigma}}
\newcommand{\tpchi}{{\bf \chi \kern -0.35em \chi}}
\newcommand{\llambda}{{\bf \lambda \kern -0.45em \lambda}}

\bibliography{plain}
\bibliography{plain}
\title{\bf Distinguishability of the symmetric states}\vspace{20mm}
\author{ M. A. Jafarizadeh$^{a}$ \thanks{E-mail:jafarizadeh@tabrizu.ac.ir}, P.Sadeghi$^{b}$
\thanks{E-mail:psadeghi@tabrizu.ac.ir}, d.Akhgar$^{a}$ \thanks{E-mail:d.akhgar@tabrizu.ac.ir}, P.Mahmoudi$^{c}$ \thanks{E-mail:p.mahmoudi@azaruniv.edu}
\\ $^a${\small Department of Theoretical Physics and Astrophysics,
University of Tabriz, Tabriz 51664, Iran.}, \\ $^b${\small Marand Faculty of Engineering, University of Tabriz, Tabriz, Iran.},\\ $^c${\small Department of Physics,  Azarbaijan Shahid Madani University. 53714-161. Tabriz. Iran}
} \pagebreak

\vspace{20mm}
\begin{document}
\maketitle \vspace{15mm}

\begin{abstract}
In this paper, the distinguishability of multipartite
geometrically uniform quantum states obtained from a single
reference state is studied in the symmetric subspace. We
specially focus our attention on the unitary transformation in a
way that the produced states remain in the symmetric subspace, so
rotation group with $J_y$ as the generator of rotation is applied.
The optimal probability and measurements are obtained for the pure
and some special mixed separable states and the results are
compared with those obtained at the previous articles for the
special cases. The results are valid for linearly dependent
states. The discrimination of these states is also investigated
using the separable measurement. We introduce appropriate
transformation to gain the optimal separable measurements
equivalent to the optimal global measurements with the same
optimal probability.
\end{abstract}

\newpage
\section{INTRODUCTION}
Discrimination of nonorthogonal quantum states is a fundamental
and important problem in quantum information theory. In
distinguishing a quantum state that belongs to the set of known
quantum states with given prior probabilities, one possibility is
to find a set of positive operator valued measure(POVM) that
maximizes the probability of correct detection, which called
minimum error
discrimination\cite{Jafarizadeh2011,Peres1991,Chitambar2013}. In
the 1970s, necessary and sufficient conditions for an optimum
measurement have been derived by Holevo, Helstrom, and Yuen
\textit{et al.} \cite{Holevo1973,Helstrom1976,Yuen1975}. However,
solving problems by means of them, except for some particular
cases, is a difficult task. Jafarizadeh \textit{et al.}
\cite{Jafarizadeh2011} presented optimality conditions, by using
Helstrom family of ensembles, which is not only powerful in
solving problems but also easy to apply. Accordingly, using this
technique, we obtain the optimal measurements.

In many discrimination problems, considerable attention is paid
to use local quantum operations and classical communication
between the components
(LOCC)\cite{Peres1991,Chitambar2013,Bennett1999}. However, using
LOCC does not have a simple mathematical structure to give
analytical optimization, and for some cases can't achieve the
optimal probability which obtained by global measurements;
therefore, obtained information reduce
\cite{Bennett1999,Chitambar2013}. In order to partially overcome
the defects, some researchers began to solve alternative problem
by separable operation, to investigate the local
distinguishability
\cite{Wootters2006,Bandyopadhyay2013,Duan2009,Xin2008}. These
operators are free of entanglement and made strict superset of
LOCC \cite{Bennett1999}.

Dimension of Hilbert space for $n$-qubit systems grows
exponentially by $n$. In order to decrease the complexity of
discrimination problem for these particles, we restrict the
problem to a set of states that possesses sufficient symmetry.
Symmetric subspace contains the states which  are invariant under
the permutations of particles. This symmetric subspace is spanned
by the $n+1$ Dicke states. Dicke states are produced and detected
experimentally \cite{Xin2014,Chiuri2010,Thiel2007}. In addition,
Dicke states are proposed for certain tasks in quantum
information theory \cite{Ivanov2010,Ivanov2010}.

In this paper, we investigate the minimum-error discrimination of
the bosonic state in many-particle spin $1/2$ systems. Selected
States have geometrically uniform (GU) symmetry in the bosonic
subspace. The set of GU states are in the form of $\{\rho
_k=U^k\rho _0(U^k)^{\dag}, k=0,1,...,m\}$, where $U$is unitary
matrices \cite{Nakahira2012}. GU states have well-known examples
such as Quadrature amplitude modulation (QAM), pulse-position
modulated ( PPM) and phase-shift-keyed( PSK) that discrimination
of them investigated extensively \cite{Kato1999,Nair2012}. We
select $\rho _0$ as pure or special mixed separable state in
symmetric subspace and as $U^k=\exp(-i2kJ_y\pi /m)$ that rotates a
spin-$j$ state by $2k\pi/m$ with respect to the $J_y$-axis. We
obtain optimal probability of correct detection and optimal
global measurement, while, results are valid for arbitrary $k$,
even for linearly dependent states. The set of pure GU states in
the symmetric subspace which are perfectly discriminated by the
obtained measurements, are identified. Also, separable form of
optimal global measurements is obtained. By Mapping the optimal
measurement from the symmetric space to entire space of
$n$-qubit, we succeed in obtain optimal separable measurements
equivalent to optimal global measurements with the same error
probability.

In Sec. II a brief review of the minimum error discrimination is
presented. In Sec. III and IV optimal detection of GU pure and
mixed states in the Symmetric subspace are investigated and the
optimal probability and the optimal global measurements are
obtained, respectively. In Sec. V appropriate transformation is
introduced  to gain the optimal separable measurements equivalent
to the optimal global measurements with the same optimal
probability, and Sec. VII is devoted to the conclusions.

\section{Minimum error discrimination }
We assume a quantum system is prepared from a collection of given
states which represented by m density operators
$\{\rho_i,\rho_i\geq0,Tr(\rho_i)=1,i=0,1...m\}$, and transmission
probability to the receiver for each of them is $p_i$ is $\sum_i
^m p_i=1$. The aim is to obtain the set of positive semidefinite
operators,  $\{ \Pi _i, \sum \Pi _i=I\}$, in the way that the
output state of operator $\Pi _i$, represent state $\rho _i$.
Therefore, the probability of correct discrimination for each
$\rho _i$  is $Tr(\rho _i \Pi _i)$. In the minimum error
approach, the set of measurement operators are looked for which
provide maximum probability of correct discrimination as follows
\begin{equation}
p_{opt}=1-p_{error}=\sum _1 ^m p_iTr(\rho _i\Pi _i).\label{eq1}
\end{equation}
The necessary and sufficient conditions of discrimination with
the maximum-success probability is
\begin{equation}
\sum _1 ^m p_i\Pi _i\rho _i-p_j\rho _j\geq0,\forall
j=1,...,m.\label{eq2}
\end{equation}
In the Ref. \cite{Jafarizadeh2011} has been demonstrated that the
necessary and sufficient conditions are equivalent to a Helstrom
family of ensembles; then a more suitable form of the conditions
of the minimum error discrimination is presented as
\begin{equation}
\textbf{M}=p_j\rho _j+(p-p_j)\tau _j,\forall j,\label{eq3}
\end{equation}
where $\textbf{M}=\sum _{i=1}^mp_i\rho _i\Pi_i$ and $\{\tau _i,
\tau _i\geq 0 \}$ is the conjugate state of $\rho _i$. Also,
eigenvector of $\tau _i$ with zero eigenvalue is proportional to
$\Pi _i$ \cite{Jafarizadeh2011},
\begin{equation}
\Pi _i\tau _i = 0.\label{eq3a}
\end{equation}
In the following two sections, the new technique is applied for
optimal detection of GU pure and mixed states in the Symmetric
subspace.

\section{Optimal detection of GU pure states in the Symmetric subspace}
In this section we derive the maximum attainable value of the
success probability in the method of the minimum error
discrimination probability for GU Symmetric states of $n$-qubit
with equal the priori probabilities.

States that are invariant under permutation of particles,
$\{P|\psi \rangle =|\psi \rangle,~ P\in S_n\}$, are called
Symmetric subspace states. For the $n$-qubit in the Hilbert space,
$\otimes _i^nH_2$, common eigenvectors  $J_z$, $J^2$ are standard
orthogonal bases and the symmetric subspace $H_s$ is indicated
with $j=n/2$. Therefore, any state in this subspace is expressed
as
\begin{equation}
|\psi _0 \rangle =\sum _{q=-\frac{n}{2}} ^{\frac{n}{2}} c_q|j,q
\rangle _z.\label{eq4}
\end{equation}
where $c_q$ is the probability amplitude. We distinguish the set
of states $\{\rho _k=U^k \rho _0(U^k) ^{\dag}\}$ , which $ \rho
_0$ is in the symmetric subspace and $U^k $ is a unitary operator.
This set is well-known as GU states. We specially are interest in
the unitary transformation which the produced states remain in
the symmetric subspace, in a way that, $\sigma _y$  and $J_y$ are
selected as generator of rotation for each qubit and generator
for $n$-qubit, respectively. Therefore, unitary transformation is
written as
\begin{equation}
U=\exp(-i\frac{\pi}{m}\sigma _y)\Rightarrow
\textbf{U}=\overbrace{U\otimes U ...\otimes U}^n
=e^{-i\frac{2\pi}{m}J _y},\label{eq5}
\end{equation}
the number of states,$m$ ,  may be equal or greater than the
dimension of $H_s$, in other words, it is not necessary to states
be linear independent .We consider states with equal initial
probability, $1/m$ , hence, from Eq.~(\ref{eq3}) for all $k$
\begin{equation}
\textbf{M}=U^k[\frac{1}{m}\rho _0+(p-\frac{1}{m})\tau
_0](U^k)^{\dag}=U^kM(U^k)^{\dag}~~k=0,1,...,m-1.\label{eq6}
\end{equation}
Thus, $\textbf{M}$ and $U^k$ commute, in addition, by
Cayley-Hamilton theorem in the subspace $j=n/2$, $\textbf{M}$ is
written as $\textbf{M}=\sum _{i=0} ^{n-1}a_iJ_y ^i$ and from
Eq.~(\ref{eq3}) one obtains
\begin{equation}
p=\sum _{i=0} ^{n}a_iTr(J_y ^i),\label{eq7}
\end{equation}
Which $p$ is Helestrom ratio and $p_{opt}<p$
\cite{Jafarizadeh2011}. Then optimization problem are given by
\begin{equation}
min ~~~~~~p=\sum _{i=0} ^{n}a_iTr(J_y ^i),\label{eq8}
\end{equation}
\begin{eqnarray}
subject~ to~~-\tau _0=-(\sum _{i=0} ^{n}a_iJ_y
^i-\frac{1}{m}|\psi _0\rangle \langle \psi _0|)\leq 0,\label{eq9}
\end{eqnarray}
and the dual problem is
\begin{equation}
max ~~~~~~g(Z_0)=\frac{1}{m} \langle \psi _0|Z_0|\psi _0
\rangle,\label{eq10}
\end{equation}
\begin{eqnarray}
subject~ to~~&& Z_0\geq 0\nonumber
\\&&Tr(J_y ^i)-Tr(Z_0J_y ^i)=0~~i=0,1,...,n-1,\label{eq11}
\end{eqnarray}
From slackness conditions  $\tau _0Z_0=0$ and  Eq.~(\ref{eq3a})
$\Pi _i$ is concluded,
\begin{eqnarray}
\Pi _0=Z_0=| z_0\rangle \langle z _0 |,\label{eq12}
\end{eqnarray}
  where $|z_0\rangle $  is expressed by eigenvector of $J_y$, $|z_0\rangle =\sum _{q=-\frac{n}{2}} ^{\frac{n}{2}}\alpha
_q|j,q\rangle _y$. Using Eq.~(\ref{eq11}), $\sum _q
q^i(1-|\alpha| _q ^2)=0$. So, for all $i$,
\begin{eqnarray}
\left[ {\begin{array}{*{20}{c}}
1&1&1& \ldots &1\\
{(\frac{n}{2})}&{(\frac{n}{2} - 1)}&{(\frac{n}{2} - 2)}& \ldots &{( - \frac{n}{2})}\\
{{{(\frac{n}{2})}^2}}&{{{(\frac{n}{2} - 1)}^2}}&{{{(\frac{n}{2} - 2)}^2}}& \ldots &{{{( - \frac{n}{2})}^2}}\\
 \vdots &{}&{}& \ldots &{}\\
{{{(\frac{n}{2})}^n}}&{}&{}& \ldots &{{{( - \frac{n}{2})}^n}}
\end{array}} \right]\left[ {\begin{array}{*{20}{c}}
{1 - \left| \alpha  \right|_{\frac{n}{2}}^2}\\
 \vdots \\
{1 - \left| \alpha  \right|_{ - \frac{n}{2}}^2}
\end{array}} \right] = 0,\label{eq12a}
\end{eqnarray}
matrix of coefficients is the same of the vandermonde matrix,
since there is no two equal rows, determinant of the matrix of
coefficients is non-zero, thus, we conclude that $|\alpha| _q
^2=1$ and $|z_0\rangle =\sum _q e^{i\theta _q} |j,q\rangle _y$.
Inserting the above result into the equation $\tau _0|z_0\rangle
=0$ , one obtains
\begin{eqnarray}
&&[\sum _{i=0} ^{n-1}a_iJ_y ^i-\frac{1}{m}|\psi_0\rangle \langle
\psi_0|]\sum _{q=-\frac{n}{2}} ^{\frac{n}{2}}e^{i\theta
_q}|j,q\rangle _y=0 \nonumber
\\
&&\sum _{q=-\frac{n}{2}} ^{\frac{n}{2}}[e^{i\theta _q}\sum _{i=0}
^{n-1}a_iq ^i-\frac{\lambda}{m}~ _y\langle j, q|\psi _0\rangle
]|j,q\rangle _y=0\nonumber
\\
&& e^{i\theta _q}\sum _{i = 0} ^{n - 1} a_iq^i = \frac{\lambda
}{m}~_y\langle j,q|\psi _0\rangle
 , \label{eq13}
\end{eqnarray}
and
\begin{eqnarray}
|\sum _{i=0} ^{n-1}a_iq ^i|=\frac{|\lambda |}{m}  |_y\langle
j,q|\psi _0\rangle |,\label{eq14}
\end{eqnarray}
where $\lambda =\langle \psi _0|z_0\rangle$.

After this, all coefficients in the initial state are real.
Equation $J_y|J,q\rangle ^* _y=-q|J,q\rangle ^* _y$, yields
\begin{eqnarray}
|_y\langle j,q|\psi _0\rangle |=|_y\langle j,-q|\psi _0\rangle
|,\label{eq15}
\end{eqnarray}
and
\begin{eqnarray}
|\sum _{i=0} ^{n}a_iq ^i|=|\sum _{i=0} ^{n}a_i(-q)
^i|.\label{eq16}
\end{eqnarray}
Hence, Helestrom ratio, $p= \sum _{i = 0} ^n a_i\sum _{q = -
\frac{n}{2}} ^{q =\frac{n}{2}} q^i$, is zero for the odd numbers
of $i$, in the Eq.~(\ref{eq8}),and only even numbers of $i$ have
non-zero terms. For $\lambda =|\lambda|e^{i\theta _\lambda}$, and
$~_y\langle j,q|\psi _0\rangle=|~_y\langle j,q|\psi _0\rangle
|e^{i\theta _{_y\langle j,q|\psi _0\rangle}}$, last term of
Eq.~(\ref{eq13}) yields $\theta _q=\theta _{\lambda}+\theta
_{_y\langle j,q|\psi _0\rangle}$, and implies:
\begin{eqnarray}
\lambda=\sum _{q=-\frac{n}{2}} ^{\frac{n}{2}}\langle \psi _0|
j,q\rangle _ye^{i\theta _q},\label{eq17}
\end{eqnarray}
so
\begin{eqnarray}
|\lambda |=\sum _{q=-\frac{n}{2}} ^{\frac{n}{2}}|~_y\langle j,q|
\psi _0\rangle |.\label{eq18}
\end{eqnarray}
 Strong duality for the optimal value of dual problem,
$p_{opt}$ , yield $p_{opt}=p$,thus
\begin{eqnarray}
p_{opt}=\sum _{i=0} ^{n}a_iTr(J_y ^i)=\sum _{q=-\frac{n}{2}}
^{\frac{n}{2}}e^{-i\theta _q}\frac{|\lambda |e^{i\theta
_{\lambda}}}{m}|~_y\langle j,q| \psi _0\rangle |e^{i\theta
_{\langle j,q| \psi _0\rangle}}=\frac{|\lambda
|^2}{m}.\label{eq19}
\end{eqnarray}
Eq.~(\ref{eq19}) is valid for all of the number of states, $m$.
The unnormalized vector $|z_0\rangle $ in the optimal measurement
operator, $\Pi _0=|z_0\rangle \langle z_0|$, is
\begin{eqnarray}
|z_0 \rangle =\sum _{q=-\frac{n}{2}} ^{\frac{n}{2}}e^{i\theta
_q}| j,q \rangle _y=\frac{1}{\sqrt{m}}\sum _{q=-\frac{n}{2}}
^{\frac{n}{2}}e^{i\theta _{~_y\langle j,q | \psi _0\rangle}}| j,q
\rangle _y. \label{eq21}
\end{eqnarray}
Therefore, one obtains the set of the optimal measurements as
$\{Z_i=U^kZ_0(U^k)^{\dag},~~k=0,1,...,m-1\}$. In the case that the
set of GU states are linearly independent, the optimal
measurements, $|z_k \rangle =U^k |z_0 \rangle $, are projective
and discrete Fourier transformation of $|z_0\rangle $ and a
projective set,
\begin{eqnarray}
\left\langle {{z_k}}| {{z_h}} \right\rangle  &=&
\frac{1}{{m}}\sum\limits_{q,\,q'} {{e^{ - i\theta
{\,_{_y\left\langle {j,q}
 | {{\psi _0}} \right\rangle }} + }}^{i\theta
{\,_{_y\left\langle {j,q'}  | {{\psi _0}} \right\rangle }}}\left(
{_y\left\langle {j,q} \right|{{\left( {{{\bf{U}}^\dag }}
\right)}^k}} \right)} \left( {{{\bf{U}}^h}{{\left| {j,q'}
\right\rangle }_y}} \right)\nonumber
\\
&=& \frac{1}{{m}}\sum\limits_{ - \frac{n}{2}}^{\frac{n}{2}}
{{e^{i\frac{{2\pi }}{m}q(k - h)}}} \nonumber
\\
&=& \frac{1}{{m}}{e^{ - i\frac{{2\pi ({n \mathord{\left/
 {\vphantom {n 2}} \right.
\kern-\nulldelimiterspace} 2}\,\, + 1)}}{m}(k -
h)}}\sum\limits_{p = 1}^{n + 1} {{e^{i\frac{{2\pi }}{m}p(k -
h)}}}  = {\delta _{k,h}}\,\,\,\,\,\forall \,\,\,\,\,n + 1 = m.
\label{eq21b}
\end{eqnarray}
This result is in agreement with pretty good measurement.

If we suppose reference state as $|\psi_0 \rangle =| j,q \rangle
_z$, it is always possible to make $\theta _{~_y\langle j,q |
\psi _0\rangle }$, constant, therefore, $|z_0 \rangle $ has a
simple form as
\begin{eqnarray}
|z _0 \rangle =\frac{1}{\sqrt{m}}\sum _{q=-\frac{n}{2}}
^{\frac{n}{2}}| j,q \rangle _y \label{eq22}
\end{eqnarray}
This result is consistent with Ref \cite{Peres1991}.

For the perfect discrimination, Eq.~(\ref{eq19}) is written as
\begin{eqnarray}
\sum _{q=-\frac{n}{2}} ^{\frac{n}{2}}e^{i\theta _q}|~_y\langle
j,q| \psi _0\rangle |=\sqrt{m}.\label{eq20}
\end{eqnarray}
For example, for generating state $|\psi _0 \rangle
=\frac{1}{\sqrt{n+1}}\sum _{q=-\frac{n}{2}}
^{\frac{n}{2}}e^{i\theta _q}| j,q \rangle _y$  , the set of GU
states, which are linearly independent, are discrete Fourier
transformation of $|\psi _0 \rangle$  and orthogonal, so
perfectly is discriminated.

\section{Optimal detection of GU mixed states in the Symmetric subspace}
For discrimination of GU mixed states in the symmetric subspace,
we select separable mixed state,
\begin{eqnarray}
\rho _0  =r| \frac{n}{2},\frac{n}{2} \rangle \langle
\frac{n}{2},\frac{n}{2}|+(1-r)| \frac{n}{2},\frac{-n}{2} \rangle
\langle \frac{n}{2},\frac{-n}{2}|, \label{eq23}
\end{eqnarray}
as the reference state. Such as the previous section, optimal
probability is obtained from the minimization of Eq.~(\ref{eq8}),
\begin{eqnarray}
(p-\frac{1}{m})\tau _0  =\sum _{i=0} ^{n-1} a_iJ_y
^i-\frac{1}{m}(r| \frac{n}{2},\frac{n}{2} \rangle \langle
\frac{n}{2},\frac{n}{2}|+(1-r)| \frac{n}{2},\frac{-n}{2} \rangle
\langle \frac{n}{2},\frac{-n}{2}|), \label{eq24 }
\end{eqnarray}
Thus, the  dual problem is written as follows:
\begin{eqnarray}
max ~~~~~~&&g(Z_0)=\frac{1}{m} (r\langle
\frac{n}{2}|Z_0|\frac{n}{2} \rangle +(1-r)\langle
\frac{-n}{2}|Z_0|\frac{-n}{2} \rangle )\nonumber
\\
subject~ to~~&&Z_0\geq 0\nonumber
\\&&Tr(J_y ^i)-Tr(Z_0J_y ^i)=0~~i=0,1,...,n-1,\label{eq26}
\end{eqnarray}
From $\tau _0|z_0\rangle =0$, and $|z_0\rangle =\sum
_{q=-\frac{n}{2}} ^{\frac{n}{2}}\alpha _q| j,q \rangle _y $,
Eq.~(\ref{eq14}) for each $q$ is written as
\begin{eqnarray}
e^{i\theta _q}\sum _{i=0} ^{n-1}a_iq ^i=\frac{r}{m}\langle j,
q|\frac{n}{2}\rangle \sum _{\acute{q}}\langle \frac{n}{2}
|j,\acute{q} \rangle e^{i\theta
_{\acute{q}}}+\frac{1-r}{m}\langle j, q|\frac{-n}{2}\rangle \sum
_{\acute{q}}\langle \frac{-n}{2} |j,\acute{q} \rangle e^{i\theta
_{\acute{q}}} , \label{eq27}
\end{eqnarray}
$e^{-i\pi J_y}|j,\frac{n}{2} \rangle _z=|j,\frac{-n}{2} \rangle
_z$ and $_y\langle j,q|j,\frac{-n}{2} \rangle =e^{i\pi q} ~
_y\langle j,q|j,\frac{n}{2} \rangle$ are concluded From
$_z\langle j,\acute{m}|e^{-i\pi J_y}|j,m \rangle _z =\delta
_{m+\acute{m},0}$. Thus, Eq.~(\ref{eq27}) is given by
\begin{eqnarray}
e^{i\theta _q}\sum _{i=0} ^{n-1}a_iq ^i=\frac{\langle j,
q|\frac{n}{2}\rangle}{m} \Omega _q, \label{eq28}
\end{eqnarray}
where  $\Omega _q =\sum _{\acute{q}}\langle \frac{n}{2}|j,
\acute{q}\rangle e^{i\theta _{\acute{q}}}[r+(1-r)e^{i\pi
(q-\acute{q})}] $. It is always possible to make  $\theta
_{y_\langle j, q|\frac{n}{2},\frac{n}{2}\rangle}=0$, for all
$q$s, and from $\Omega _q=\Omega _{q+2}$, is concluded which
$exp(i\theta _q)$ takes only two different values. Without loss
the generality, phase of $|j,-j+2s\rangle _y$ and
$|j,-j+(2s+1)\rangle _y$ for $s=0,1,...$ is given $1$ and
$\exp(i\theta )$, respectively. Therefore,
\begin{eqnarray}
\Omega _q &=&\sum _s\langle \frac{n}{2}|j,-j+2s \rangle
[r+(1-r)e^{i\pi (q+\frac{n}{2})}e^{-i\pi 2s}]\nonumber
\\
&+& e^{i\theta}\sum _s\langle \frac{n}{2}|j,-j+(2s+1) \rangle
[r+(1-r)e^{i\pi (q+\frac{n}{2})}e^{-i\pi (2s+1)}]\nonumber
\\
&=& (r+(1-r)e^{i\pi (q+\frac{n}{2})})A+e^{i\theta}(r+(1-r)e^{i\pi
(q+\frac{n}{2})})B , \label{eq29}
\end{eqnarray}
where $\sum _s\langle \frac{n}{2}|j,-j+2s \rangle =A$, $\sum
_s\langle \frac{n}{2}|j,-j+(2s+1) \rangle =B$, and $\Omega
_{q=-j+2s}=A+e^{i\theta}(2r-1)B$, $e^{-i\theta}\Omega
_{q=-j+(2s+1)}=(2r-1)e^{-i\theta}A+B$.\\
According to obtained equations, optimal success probability is
given by
\begin{eqnarray}
p_{opt}&=&\sum _{q=-\frac{n}{2}} ^{q=\frac{n}{2}} e^{-i\theta
_q}\frac{\langle j, q|\frac{n}{2}\rangle}{m} \Omega
_q=\frac{\Omega _{q=-j+2s}}{m}A +\frac{e^{-i\theta }\Omega
_{q=-j+(2s+1)}}{m}B\nonumber \\
&=&\frac{1}{m}[(\sum _q\langle j,
q|\frac{n}{2}\rangle)^2+2AB(\cos(\theta)(2r-1)-1)], \label{eq30}
\end{eqnarray}
From Eq.~(\ref{eq30}), $\cos(\theta)=1$  and $\cos(\theta)=-1$ for
$r>\frac{1}{2}$ and  $r<\frac{1}{2}$ , respectively. For the
systems with odd number of qubit, $A=B$. Therefore, $p_{opt}$ is
simplified to the following form
\begin{eqnarray}
p_{opt}=\frac{\cos(\theta)(2r-1)-1}{2m}(\sum _q\langle j,
q|\frac{n}{2}\rangle)^2. \label{eq31}
\end{eqnarray}
As a special case, for mixed three-qubit states, $p_{opt}$ becomes
\begin{eqnarray}
p_{opt}=\frac{1}{3}(1+|2r-1|). \label{eq32}
\end{eqnarray}
This is in agreement with the results in the Ref.
\cite{Jafarizadeh2010}.

Like the pure states for the linearly independent states, $m=n+1$,
the optimal measurements are discrete Fourier transformations of
$|z_0\rangle $ and projective.

\section{Finding optimal separable measurements}
In the Majorana representation any symmetric state of $n$-qubit,
$|\phi_S\rangle$,  which is invariant under the
permutation, is uniquely made from the sum of all permutations of
$n$ single qubit state as
\begin{eqnarray}
|\phi_s \rangle =\frac{1}{\sqrt{k}}\sum _{g=S_n}g|\varphi
_1\rangle |\varphi _2\rangle ...|\varphi _n\rangle, \label{eq33}
\end{eqnarray}
which $k$ is the normalization factor. The vector of $|\varphi
_i\rangle $ is made from the roots of the following function
\begin{eqnarray}
\Phi (t) =\sum _{k=0} ^n (-1)^k \left(
\begin{array}{l}
n\\
k
\end{array}
\right)^{\frac{1}{2}}a_kt^k, \label{eq34}
\end{eqnarray}
where $t=e^{i\varphi}\tan (\theta)$, $|\varphi _i \rangle =\cos
\theta |0\rangle +e^{i\varphi}\sin \theta |1\rangle$ and $a_k$ is
expansion coefficient of $|\phi_S \rangle$ in the Dick
basis.

Since $ |z_0\rangle$  is in the symmetric subspace thus it is
always possible to write $ |z_0\rangle$  in the Majorana
representation as
\begin{eqnarray}
|z _0\rangle =\frac{1}{\sqrt{k}}\sum _{g=S_n}g|\varphi _1\rangle
|\varphi _2\rangle ...|\varphi _n\rangle = \frac{1}{\sqrt{k}}\sum
_{g\in S_n}g|\varphi _0\rangle, \label{eq35}
\end{eqnarray}
the term of $|\varphi _1\rangle |\varphi _2\rangle ...|\varphi
_n\rangle$ is inserted by $|\varphi _0\rangle$.

Similar to the general furrier transformation, we introduce the
following map on quantum state $|\varphi \rangle$
\begin{eqnarray}
|\tau _{ij} ^{\rho}\rangle =\sqrt{\frac{d_{\rho}}{n!}}\sum _{g\in
 S_n}\rho _{ij}(g)g|\varphi \rangle, \label{eq37}
\end{eqnarray}
where $S_n$ is Symmetric group and $\rho (g)$ is an irreducible
representation of $S_n$ with the dimension of $d_{\rho }$. From
Eq.~(\ref{eq19}), Majorana representation is equivalent, up to a
constant,  to transformed form of $|\varphi _0 \rangle$, by the
trivial representation, $|\tau ^0\rangle =\frac{1}{\sqrt{n!}}\sum
_{g\in
 S_n}(g|\varphi _0 \rangle )$.

Here, we prove one of the important properties of this map which
$\left| {{\tau ^0}} \right\rangle $ and each symmetric
state,$|\psi _0 \rangle$, is orthogonal to the transformed form
of non-trivial representation of $S_n$. By multiplying of
$\langle \psi _0|$, into the Eq.~(\ref{eq37}) one obtains
\begin{eqnarray}
\langle \tau ^0|\tau ^{\rho }_{ij}\rangle &=&\langle \tau
^0|\sqrt{\frac{d_{\rho}}{n!}}\sum _{g\in S_n}\rho _{ij}
(g)g|\varphi _0 \rangle\nonumber
\\&=&\sqrt{\frac{d_{\rho}}{n!}}\sum _{g\in S_n}\rho _{ij}(g)\langle \tau ^0|g|\varphi _0 \rangle\nonumber
\\&=&\langle \tau ^0|\varphi _0 \rangle\sqrt{\frac{d_{\rho}}{n!}}\sum _{g\in S_n}\rho _{ij}(g). \label{eq41}
\end{eqnarray}
In this step, we show that $\sum _{g\in S_n}\rho (g)$ equal to
zero. If $\sum _{g\in S_n}\rho (g)=A$ then for all $\acute{g}\in
G$ is obtained
\begin{eqnarray}
\rho (\acute{g})A=A\rho (\acute{g}), \label{eq42}
\end{eqnarray}
and from the shor lemma for an irreducible representation
\begin{eqnarray}
\sum _{g\in S_n}\rho (g)=\lambda I_{d_{\rho}}. \label{eq43}
\end{eqnarray}
Taking the trace of both sides leads to
\begin{eqnarray}
\sum _{g\in S_n}\chi _{\rho} (g)=\lambda {d_{\rho}}, \label{eq44}
\end{eqnarray}
where $\chi _{\rho} (g)$ is the character of group elements. For
the finite group $\sum _{g\in S_n}\chi _{\rho} (g)\chi
_{\acute{\rho}} ^* (g)=0$, and for the trivial representation
$\chi _{\acute{\rho}} ^* (g)=1$. Therefore, one obtains $\lambda
=0$, and consequently $\sum _{g\in S_n}\rho (g)=0$.This indicates
that $\sum _{g\in S_n}\rho _{ij} (g)=0$, thus,
\begin{eqnarray}
\langle \tau ^0|\tau _{i\,j}^\rho \rangle  = 0~~~~\forall \rho
\neq 1.\label{eq45a}
\end{eqnarray}
So, $\left| {\tau _{i\,j}^\rho } \right\rangle $ by non-trivial
representations, is not in the symmetric space. In fact from the
Schur-Weyl duality theorem \cite{Schur1927} the Hilbert space of
$n$-qubit, $H_2^{ \otimes n}$, expressed in term of subspaces
which are invariant under irreducible representation (irrep) of
${S_n},{U_2}$, i.e,
\begin{eqnarray}
H_2^{ \otimes n} = \sum \bigoplus {{H_{irrep
\,\,{S_n},{U_2}}}}.\label{eq45b}
\end{eqnarray}
Then, $\left|
{{\tau ^0}} \right\rangle $, Belongs to the Symmetric Hilbert
subspace and from ~(\ref{eq45a}) all $\left| {\tau _{i\,j}^\rho }
\right\rangle $, which $\rho \ne 1$, are the states in the other
subspaces.

In following, in order to find optimum separable operation, the
orthogonal terms are added to the trivial representation.
\begin{eqnarray}
 \sum _{ij}\sum _{\rho}|\tau ^{\rho }_{ij}\rangle \langle \tau ^{\rho }_{ij}|
  &=& \sum _{\acute{g},g\in S_n}\sum _{ij,\rho}\frac{d_{\rho}}{n!}\rho _{ij}
 (g)\rho ^* _{ij}(\acute{g})g|\varphi_0\rangle \langle \varphi_0|\acute{g}\nonumber
 \\ &=& \sum _{\acute{g},g\in S_n}\sum _{\rho}\frac{d_{\rho}}{n!}Tr(\rho
 (g)\rho ^{\dag} (\acute{g}))g|\varphi_0\rangle \langle \varphi_0|\acute{g}\nonumber
 \\ &=& \sum _{\acute{g},g\in S_n}\sum _{\rho}\frac{d_{\rho}}{n!}Tr(\rho
 (g\acute{g}^{-1}))g|\varphi_0\rangle \langle \varphi_0|\acute{g}\nonumber
 \\ &=& \sum _{\acute{g},g\in S_n}[\sum _{\rho}\frac{d_{\rho}}{n !}\chi _{\rho}
 (g\acute{g}^{-1})]g|\varphi_0\rangle \langle \varphi_0|\acute{g}\nonumber
 \\ &=& \sum _{\acute{g},g\in S_n}\delta _{g,\acute{g}}g|\varphi_0\rangle \langle
  \varphi_0|\acute{g}\nonumber
 \\ &=& \sum _{g\in S_n}g|\varphi_0\rangle \langle
 \varphi_0|g\nonumber
 \\ &=& \sum _{g\in S_n}g[|\varphi_1\rangle |\varphi_2\rangle
 ...|\varphi_n\rangle \langle
 \varphi_1| \langle
 \varphi_2|...\langle
 \varphi_n|]g \nonumber
 \\ &=& \left( {\left| {{\varphi _1}} \right\rangle \left\langle
{{\varphi _1}} \right| \otimes \left| {{\varphi _2}}
\right\rangle \left\langle {{\varphi _2}} \right| \otimes
...\left| {{\varphi _n}} \right\rangle \left\langle {{\varphi
_n}} \right|} \right) + \left( {\left| {{\varphi _2}}
\right\rangle \left\langle {{\varphi _2}} \right| \otimes \left|
{{\varphi _1}} \right\rangle \left\langle {{\varphi _1}} \right|
\otimes ...\left| {{\varphi _n}} \right\rangle \left\langle
{{\varphi _n}} \right|} \right) + ... \nonumber
 \\ &+& \left( {\left| {{\varphi _n}} \right\rangle \left\langle {{\varphi _n}} \right| \otimes \left| {{\varphi _1}} \right\rangle \left\langle {{\varphi _1}} \right| \otimes ...} \right) + ....
 .\label{eq45}
\end{eqnarray}
According to the last term in the Eq.~(\ref{eq45}) is yielded
each term of last summation is separable, and the probability of
correct discrimination pure state, in the symmetric subspace, by
this local operator is
\begin{eqnarray}
p^{local} _{opt}& =& \frac{1}{m}Tr( c\sum _{ij} \sum _{\sigma} |
\tau _{ij}^{\sigma } \rangle \langle \tau _{ij}^{\sigma } | \rho
_0 ) =\frac{1}{m}Tr( \sqrt{c}| \tau ^0 \rangle \langle \tau ^0 |
 \sqrt{c}\rho _0 )+ \frac{c}{m}\sum _{ij} \sum _{\sigma \neq 1} Tr( |\tau _{ij}
^{\sigma } \rangle \langle
\tau _{ij} ^{\sigma } | \rho _0)\nonumber \\
 &=& p_{opt} ^{global} + c\sum _{ij}  \sum _{\sigma \neq 1} |\langle \tau _{ij} ^{\sigma
}| \psi _0 \rangle | = p_{opt}^{global}. \label{eq45c}
\end{eqnarray}
where, $\left| {{z_0}} \right\rangle  = \sqrt c \left| {{\tau
^0}} \right\rangle $.Therefore, the optimum separable operation
equivalent to the optimum global operation is achieved.

\section{CONCLUSION}
The discrimination of GU states in the symmetric subspace of
$n$-qubit particles is investigated. In this subspace for the
general pure states, the optimal probability and the optimal
global measurements are obtained by the presented method in the
Ref. \cite{Jafarizadeh2011}. Also, as a special case, the mixed
reference states which Included the convex combination of up and
down states are studied and the results are consistent with the
results in the Ref. \cite{Peres1991} and \cite{Jafarizadeh2010}.
The following, we introduce a mapping to gain the optimal
separable measurements equivalent to the optimal global
measurements with the same optimal probability.  The achieved
results are always valid for make of the separable operations
which are in the symmetric subspace.  We expect to the same
results are expressed in the other subspaces and this paper
offers a good starting point. To make the states which are
orthogonal to symmetric subspace, we use Majorana
representation,however, this representation is not valid for
general case of qudit state, so, finding of the appropriate map to
make the separable measurement from each symmetric measurement,
is difficult and under investigation.



\end{document}